\documentstyle[nato,epsfig]{crckapb}

\begin{opening}

\title{High-energy Neutrino Astronomy: Science and First Results$^*$}
\author{Francis Halzen}
\institute{Department of Physics, University of Wisconsin,\\
1150 University Avenue,
Madison, WI 53706}

\end{opening}

\runningtitle{High-Energy Neutrino Astronomy}

\def\includegraphics[#1]#2{\epsfig{file=#2,#1}}

\newcommand{\diffunit}{${\rm s^{-1}\,cm^{-2}\,sr^{-1}\,GeV}$}
\newcommand{\pointunit}{${\rm s^{-1}\,cm^{-2}\,GeV}$}
\newcommand{\Nch}{$N_{{\rm ch}}$}

\def\gsim{\mathrel{\raise.3ex\hbox{$>$\kern-.75em\lower1ex\hbox{$\sim$}}}}

\begin{document}

\addtolength{\oddsidemargin}{1.5cm}
\evensidemargin\oddsidemargin
\addtolength{\topmargin}{,75cm}

\vglue-2.5in

\hbox to \hsize{{\bf University of Wisconsin - Madison}
\hfill\vtop{\normalsize
\hbox{\bf MADPH-02-1316}
\hbox{December 2002}
\hbox{\hfil}}}

\vglue2.25in

\renewcommand{\thefootnote}{\fnsymbol{footnote}}
\setcounter{footnote}{1}
\footnotetext{
Talk presented at the 9th Course of Astrofundamental Physics, International School of Astrophysics D.~Chalonge, Palermo, Sicily, Sept.~2002.}

\begin{abstract}
{\small
We introduce neutrino astronomy starting from the observational fact that Nature
accelerates protons and photons to energies in excess of $10^{20}$ and
$10^{13}$\,eV, respectively. Although the discovery of cosmic rays dates
back a century, we do not know how and where they are
accelerated. We review the observations as well as speculations about the
sources. Among these gamma ray bursts and active galaxies represent
well-motivated speculations because these are also the sources of the
highest energy gamma rays, with emission observed up to 20\,TeV,
possibly higher.

We discuss why cosmic accelerators are expected to be cosmic beam
dumps producing neutrino beams associated with the highest
energy cosmic rays. Cosmic ray sources may produce neutrinos from MeV to
EeV energy by a variety of mechanisms. The important conclusion is that,
independently of the specific blueprint of the source, it takes a
kilometer-scale neutrino observatory to detect the neutrino beam
associated with the highest energy cosmic rays and gamma rays. The
technology for commissioning such instrument has been established by the AMANDA detector at the South Pole. We review its performance and, with several thousand neutrinos collected, its first scientific results.
\par}
\end{abstract}

\section{The Highest Energy Particles: Cosmic Rays, Photons and
Neutrinos}

\subsection{The New Astronomy}

While conventional astronomy spans 60 octaves in photon frequency, from
$10^4$\,cm radio-waves to $10^{-14}$\,cm  gamma rays of GeV energy,
successful efforts are underway to probe the Universe at yet smaller
wavelengths  and larger photon energies; see Fig.\,1. Gamma rays,
gravitational waves, neutrinos and very high-energy protons are explored
as astronomical messengers. As exemplified time and again, the
development of novel ways of looking into space invariably results in
the discovery of unanticipated phenomena. As is the case with new
accelerators,
observing the predicted is somewhat disappointing.

\begin{figure}[t!]
\centering\leavevmode
\includegraphics[width=4in]{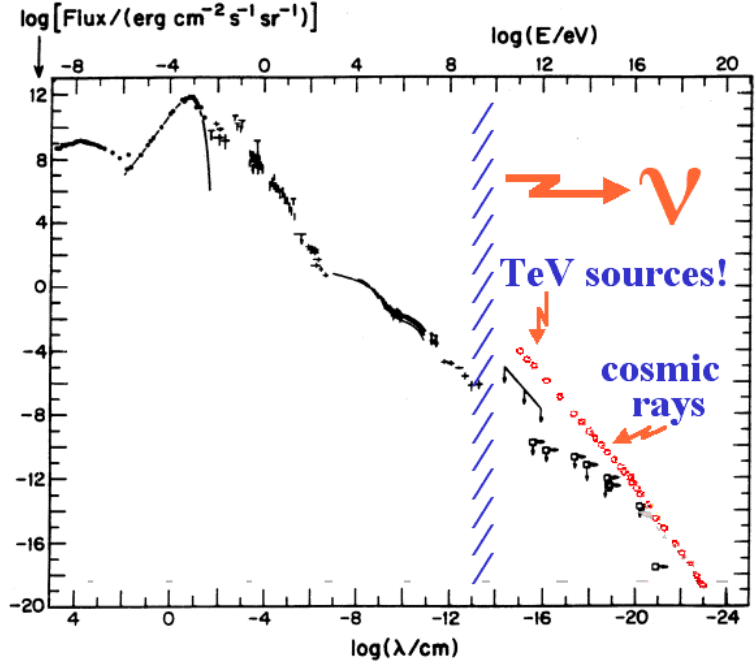}
\caption{The diffuse flux of photons in the Universe, from radio waves to
GeV-photons. Above tens of GeV, only limits are reported although
individual sources emitting TeV gamma rays have been identified. Above
GeV
energy, cosmic rays dominate the spectrum.}
\label{one}
\end{figure}

Why pursue high-energy astronomy with neutrinos or protons despite the
considerable instrumental challenges? A mundane
reason is that the
Universe is not transparent to photons of TeV energy and above
(in ascending factors of $10^3$, units are: GeV/TeV/PeV/EeV/ZeV ). For
instance, a PeV energy photon cannot deliver information from a
source
at the edge of our own galaxy because it will annihilate into an
electron pair in an encounter with a 2.7 Kelvin microwave
photon before reaching our telescope. Only neutrinos can reach us
without attenuation from the edge of the Universe at all energies.

At EeV energies, proton astronomy may be possible. Above 50\,EeV the
arrival directions of electrically charged cosmic rays are
no longer scrambled by the ambient magnetic field of our own
galaxy.    
They point back to their sources with an accuracy determined by their
gyroradius in the intergalactic magnetic field $B$:
\begin{equation}
{\theta\over0.1^\circ} \cong { \left( d\over 1{\rm\ Mpc} \right)
\left( B\over 10^{-9}{\rm\,G} \right) \over \left( E\over
3\times10^{20}\rm\, eV\right) }\,,
\end{equation}
where $d$ is the distance to the source. Speculations on the strength of
the inter-galactic magnetic field
range from $10^{-7}$ to $10^{-12}$~Gauss in the local cluster. For a
distance of 100~Mpc,
the resolution may therefore be anywhere from sub-degree to
nonexistent. Proton astronomy should
be possible at the very highest energies; it may also provide indirect
information on
intergalactic magnetic fields. Determining their strength by conventional astronomical means has been challenging.

\subsection{The Highest Energy Cosmic Rays: Facts}

In October 1991, the Fly's Eye cosmic ray detector recorded an event
of energy $3.0\pm^{0.36}_{0.54}\times 10^{20}$\,eV \cite{flyes}.
This event, together with an event recorded by the Yakutsk air shower
array in May 1989 \cite{yakutsk}, of estimated energy \mbox{$\sim
2\times10^{20}$\,eV}, constituted at the time the highest
energies recorded. Their energy corresponds to a center of
mass energy of the order of 700~TeV or $\sim 50$ Joules, almost 50
times the energy of the Large Hadron Collider (LHC). In fact, all
active 
experiments \cite{web} have detected
cosmic rays in the vicinity of 100~EeV since their initial discovery by
the
Haverah Park air shower array \cite{WatsonZas}. The AGASA air shower
array in Japan\cite{agasa} has now accumulated an impressive 10
events with energy in excess of $10^{20}$\,eV \cite{ICRC}.

The accuracy of the energy resolution of these experiments is a critical
issue. With a particle flux of order 1 event per km$^2$ per
century, these events are studied by using the earth's
atmosphere as a particle detector. The experimental signature of an
extremely high-energy cosmic particle is
a shower initiated by the particle. The primary particle creates an
electromagnetic and
hadronic cascade.  The electromagnetic shower grows to a shower maximum,
and is subsequently absorbed by the
atmosphere. The shower can be observed by: i)
sampling the electromagnetic and hadronic components when they reach
the ground
with an array of particle detectors such as scintillators, ii)
detecting the fluorescent
light emitted by atmospheric nitrogen excited by the passage of the
shower particles, iii) detecting the Cerenkov light emitted by the
large number
of particles at shower maximum, and iv)~detecting muons and neutrinos produced in the hadronic component of the air shower.

The bottom line on energy measurement is that, at this time, several
experiments using the first two techniques agree on the energy of
EeV-showers within a resolution of $\sim$\,25\%. Additionally, there is a
systematic error of order 10\% associated with the modeling of the
showers. All techniques are indeed subject to the ambiguity of particle
simulations that involve physics beyond the LHC. If the final outcome
turns out to be an erroneous inference of the energy of the shower
because
of new physics associated with particle interactions at the
$\Lambda_{\mathrm{QCD}}$ scale, we will have to contemplate that
discovery
instead.

The premier experiments, HiRes and AGASA, agree that cosmic rays with
energy in excess of 10\,EeV are not galactic in origin and that
their spectrum extends beyond 100\,EeV. They disagree on almost
everything else. The AGASA experiment claims evidence that the highest
energy cosmic rays come
from point sources, and
that they are mostly heavy nuclei. The HiRes data does not support
this. Because of low statistics, interpreting the measured fluxes as
a function of energy is like reading tea leaves; one cannot help however
reading different messages in the spectra; see Fig.\,2 and Fig.\,3.

\begin{figure}[b!]
\centering\leavevmode
\includegraphics[width=3.7in]{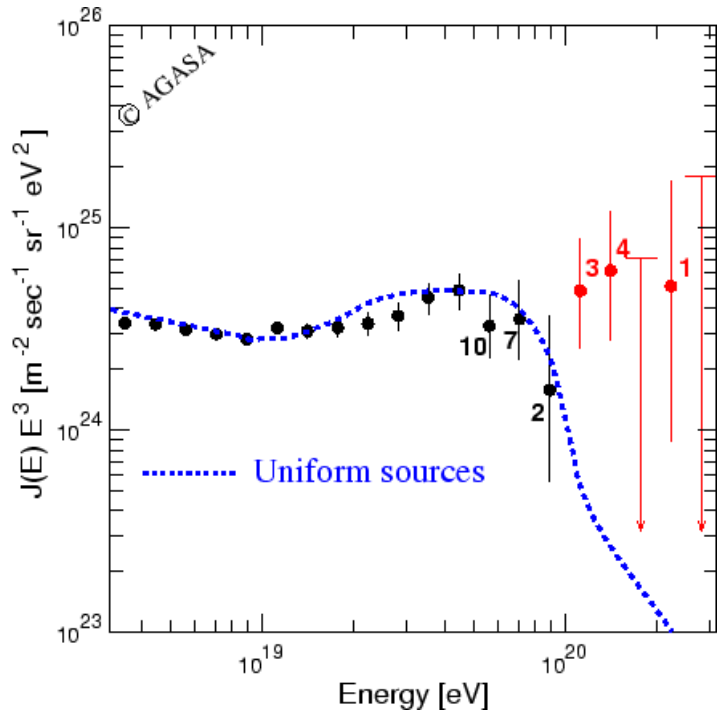}
\caption{The cosmic ray spectrum peaks in the vicinity of 1\,GeV and has
features near $10^{15}$ and $10^{19}$\,eV referred to as the ``knee" and
``ankle" in the spectrum, respectively. Shown is the flux of the highest
energy cosmic rays near and beyond the ankle measured by the AGASA
experiment.  Note that the flux is multiplied by $E^3$.}
\label{two}
\end{figure}

\begin{figure}[t!]
\centering\leavevmode
\includegraphics[width=4in]{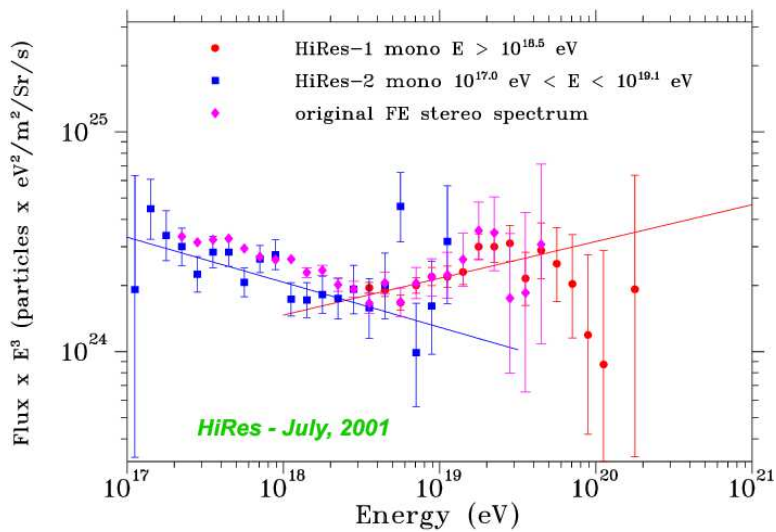}
\caption{As in Fig.\,2, results from the HiRes experiment.}
\label{three}
\end{figure}

\subsection{The Highest Energy Cosmic Rays: Fancy}

\subsubsection{Acceleration to $>100$ EeV?}

It is sensible to assume that, in order to accelerate a proton to
energy $E$ in a magnetic field $B$, the size $R$ of the accelerator
must be larger than the gyroradius of the particle:
\begin{equation}
R > R_{\rm gyro} = {E\over B}\,.
\end{equation}
That is, the accelerating magnetic field must contain the particle
orbit. This condition yields a maximum energy
\begin{equation}
E \sim \gamma BR
\end{equation}
by dimensional analysis and nothing more. The $\gamma$-factor has
been included to allow for the possibility that we may not be at rest
in the frame of the cosmic accelerator resulting in the
observation of boosted particle energies.
Theorists' imagination regarding the accelerators has been limited to
dense regions where exceptional gravitational forces
create relativistic particle flows: the dense cores of exploding
stars, inflows on supermassive black holes at the centers of active
galaxies, annihilating black holes or neutron stars. All
speculations involve collapsed objects and we can therefore replace $R$
by the Schwartzschild radius
\begin{equation}
R \sim GM/c^2
\end{equation}
to obtain
\begin{equation}
E \propto \gamma BM \,.
\end{equation}
Given the microgauss magnetic field of our galaxy, no structures are
large or massive enough to reach the energies of the highest energy
cosmic rays. Dimensional analysis therefore limits their sources to
extragalactic objects; a few common speculations are listed in
Table\,1.

\begin{table}[h]
\caption{Requirements to generate the highest energy cosmic rays in
astrophysical sources.}
\centering\leavevmode
\begin{tabular}{llll}
\hline
\multicolumn{4}{c}{Conditions with $E \sim 10\rm\ EeV$}\\
\hline
$\bullet$\ Quasars& $\gamma\cong 1$& $B\cong 10^3$ G& $M\cong 10^9
M_{\rm sun}$\\
$\bullet$\ Blazars& $\gamma\gsim 10$& $B\cong 10^3$ G& $M\cong 10^9
M_{\rm sun}$\\
$\bullet$\ \parbox[t]{1.0in}{Neutron Stars\\ Black Holes\\$\vdots$}&
$\gamma\cong 1$& $B\cong 10^{12}$ G& $M\cong M_{\rm sun}$\\
$\bullet$\ GRB& $\gamma\gsim 10^2$& $B\cong 10^{12}$ G& $M\cong M_{\rm
sun}$\\
\hline
\end{tabular}
\end{table}

Nearby active galactic nuclei, distant by $\sim100$~Mpc and
powered by a billion solar mass black holes, are candidates.
With kilogauss fields, we reach 100\,EeV. The jets (blazars) emitted by
the
central black hole could reach similar energies in accelerating
substructures (blobs) boosted in our direction by Lorentz factors of 10, possibly higher. The neutron star or black hole remnant of a
collapsing
supermassive star could support magnetic fields of $10^{12}$\,Gauss,
possibly larger. Highly relativistic shocks with $\gamma > 10^2$
emanating
from the
collapsed black hole could be the origin of gamma ray bursts and,
possibly, the source of the highest energy cosmic rays.

The above speculations are reinforced by the fact that the sources
listed are also the sources of the highest energy gamma rays
observed. At this point, however, a
reality check is in order. The above
dimensional analysis applies to the Fermilab accelerator: 10 kilogauss
fields over several kilometers corresponds to 1\,TeV. The argument holds
because, with optimized design and perfect alignment of magnets, the
accelerator
reaches efficiencies matching the dimensional limit. It is highly
questionable that nature can achieve this feat. Theorists can
imagine acceleration in shocks with an efficiency of perhaps 10\%.

The astrophysics of accelerating particles to Joule energies is so
daunting that many believe that
cosmic rays are not the beams of cosmic accelerators but the decay
products of remnants from the early Universe, such as topological
defects associated with a Grand Unified Theory (GUT) phase transition.

\subsubsection{Are Cosmic Rays Really Protons: the GZK Cutoff?}

All experimental signatures agree on the particle nature of the
cosmic rays --- they look like protons or, possibly, nuclei. We
mentioned at the beginning of this article that the Universe is
opaque to photons with energy in excess of tens of TeV because they
annihilate into electron pairs in interactions with the infrared photon background.
Protons also interact with background
light, predominantly by photoproduction of the $\Delta$-resonance,
i.e.\ $p + \gamma_{CMB} \rightarrow \Delta \rightarrow \pi + p$ above
a threshold energy $E_p$ of about 50\,EeV given by:
\begin{equation}
2E_p\epsilon > \left(m_\Delta^2 - m_p^2\right) \,.
\label{eq:threshold}
\end{equation}
The major source of proton energy loss is photoproduction of pions on
a target of cosmic microwave photons of energy $\epsilon$. The
Universe is, therefore, also opaque to the highest energy cosmic rays,
with an absorption length~of
\begin{eqnarray}
\lambda_{\gamma p} &=& (n_{\rm CMB} \, \sigma_{p+\gamma_{\rm
CMB}})^{-1}\cong10\rm\ Mpc
\end{eqnarray}
when their energy exceeds 50\,EeV. This
so-called GZK cutoff establishes a universal upper limit on
the energy of the cosmic rays. The cutoff is robust,
depending only on two known numbers: $n_{\mathrm{CMB}} = 400\rm\,cm^{-3}$
and
$\sigma_{p+\gamma_{\mathrm{CMB}}} = 10^{-28}\rm\,cm^2$.

Cosmic rays do reach us with energies exceeding 100\,EeV. This presents
us with three options: i) the protons are
accelerated in nearby sources, ii)~they do reach us from distant sources
which accelerate them to even higher energies than we observe, thus
exacerbating the acceleration problem, or iii) the highest energy cosmic
rays are not protons.

The first possibility raises the considerable challenge of finding an appropriate
accelerator in our
local galactic cluster. It is not impossible that all cosmic rays are
produced by the active galaxy M87, or by a nearby gamma ray burst
which exploded a few hundred years ago.

\looseness=-1
Stecker \cite{stecker2} has speculated that the highest energy cosmic
rays may be Fe nuclei with a delayed GZK cutoff. The details are
complicated but the relevant quantity in the problem is $\gamma=E/AM$,
where $A$ is the atomic number and $M$ the nucleon mass. For a fixed
observed energy, the smallest boost towards GZK threshold is associated
with the largest atomic mass, i.e.~Fe.

\subsubsection{Could Cosmic Rays be Photons or Neutrinos?}

The above question naturally emerges in the context of models where the
highest energy cosmic rays are the decay products of remnants or
topological structures created in the early universe with typical energy
scale of order $10^{24}$\,eV. In these scenarios the highest energy
cosmic rays are predominantly photons. A topological defect will suffer
a chain decay into Grand Unified Theory (GUT) particles X and Y, that
subsequently
decay to familiar weak bosons, leptons and quark or gluon jets. Cosmic
rays are, therefore, predominately the fragmentation products of these
jets. We know from accelerator studies that, among the fragmentation
products of jets, neutral pions (decaying into photons) dominate, in
number, protons by close to two orders of magnitude. Therefore, if the
decay of topological defects is the source of the highest energy cosmic
rays, they must be photons. This is a problem because there is
compelling evidence that the highest energy cosmic rays are not photons:

\begin{enumerate}
\item The highest energy event observed by Fly's Eye is not likely to
be a photon \cite{vazquez}.  A photon of 300\,EeV will interact with the
magnetic field of the earth far above the atmosphere and disintegrate
into lower energy cascades --- roughly ten at this particular energy.
The detector subsequently collects light produced by the fluorescence of
atmospheric nitrogen along the path of the high-energy showers
traversing the atmosphere. The atmospheric shower profile of a 300\,EeV
photon after fragmentation in the earthÕs magnetic field, is shown in
Fig.\,4. It disagrees with the data. The observed shower
profile does fit that of a primary proton, and,
possibly, that of a nucleus. The shower profile information is
sufficient,
however, to conclude that the
event is unlikely to be of photon origin.

\item The same conclusion is
reached for the Yakutsk event that is characterized by a huge number
of secondary muons, inconsistent with a pure electromagnetic cascade
initiated by a gamma ray.

\item The AGASA collaboration claims evidence
for ``point" sources above 10\,EeV. The arrival directions are
however smeared out in a way consistent with primaries deflected by
the galactic magnetic field. Again, this indicates charged primaries
and excludes photons.

\item Finally, a recent reanalysis of the Haverah Park disfavors photon
origin of the primaries \cite{WatsonZas}.

\end{enumerate}

\begin{figure}[t]
\centering\leavevmode
\includegraphics[height=3.5in]{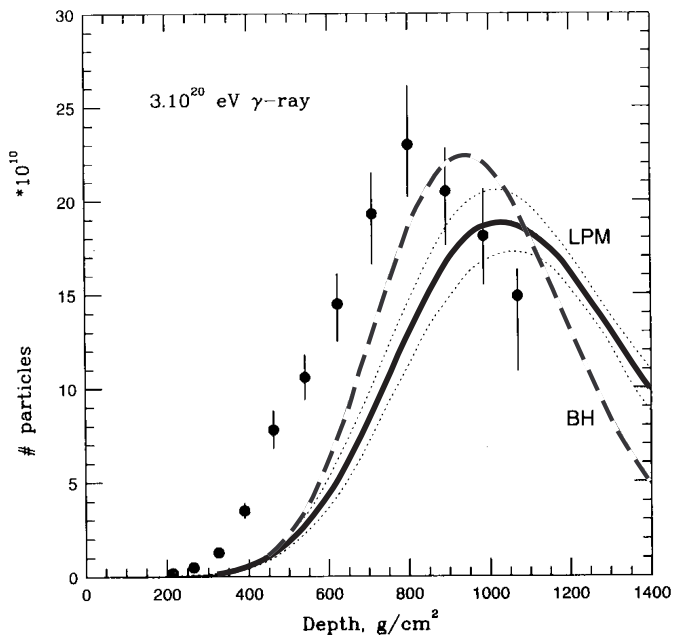}
\vskip-2ex

\caption{The composite atmospheric shower profile of a $3\times
10^{20}$\,eV gamma ray shower calculated with Landau-Pomeranchuk-Migdal
(dashed) and
Bethe-Heitler (solid)
electromagnetic cross sections. The central line shows the average shower
profile and the upper and lower lines show 1~$\sigma$ deviations --- not
visible for the BH case, where lines overlap. The experimental shower
profile
is shown with the data points. It does not fit the profile of a photon
shower.}
\label{four}
\end{figure}

Neutrino primaries are definitely ruled out. Standard model neutrino
physics is understood, even for EeV energy. The average $x$ of
the parton mediating the neutrino interaction is of
order $x \sim \sqrt{M_W^2/s} \sim 10^{-6}$ so that the perturbative
result for the neutrino-nucleus cross section is calculable from measured
HERA structure functions. Because $Q^2\sim{M_W}^2$, even at 100\,EeV a
reliable value of the cross section can be obtained based on QCD-inspired
extrapolations of the structure function. The neutrino cross section is
known to better than an order of magnitude. It falls 5 orders of
magnitude
short of the strong cross sections required to make a neutrino interact
in
the upper atmosphere to create an air shower.

Could EeV neutrinos be strongly interacting because of new physics?
In theories with TeV-scale gravity, one can imagine that graviton
exchange dominates all interactions thus erasing the difference
between
quarks and
neutrinos at the energies under consideration. The actual models
performing this feat require a fast turn-on of the cross
section with energy that (arguably) violates S-wave unitarity
\cite{han}.

We have exhausted the possibilities.  Neutrons, muons and other
candidate primaries one may think of are unstable. EeV neutrons
barely live long enough to reach us from sources at the edge of our
galaxy.

\section{A Three Prong Assault on the Cosmic Ray Puzzle}

We conclude that, where the highest energy cosmic rays are concerned,
both
the
accelerator mechanism and the particle physics are enigmatic. The mystery
has inspired a worldwide effort to tackle the
problem with novel experimentation including air shower arrays covering
an area of several times $10^3$ square kilometers\cite{watson} and
arrays of multiple air Cerenkov telescopes\cite{volk}. We here discuss
kilometer-scale neutrino observatories. While these
have additional missions such as the search for dark matter,
their observations are expected to have an impact on cosmic ray physics.

We anticipate indeed that secondary photons and neutrinos are associated
with the highest energy cosmic rays; see Fig.\,5. The cartoon
draws our attention to the fact that cosmic accelerators are also cosmic
beam dumps that produce secondary
photon and neutrino beams.
Accelerating particles to TeV energy and above requires relativistic,
massive bulk flows. These are likely to originate from the
exceptional gravitational forces associated with black holes or
neutron stars. Accelerated particles therefore pass through intense
radiation fields or dense
clouds of gas surrounding the black hole leading to the production of
secondary pions. These subsequently decay into photons and neutrinos
that accompany the primary cosmic ray beam. Examples of targets for muon production include the external photon clouds and the UV radiation field that
surrounds the central black hole of active galaxies, or the matter
falling into the collapsed core of a dying supermassive star producing a
gamma ray burst.
The target material, whether a gas of particles or of photons,
is likely to be sufficiently tenuous for the primary proton beam
and the secondary photon beam to be only partially attenuated. However,
shrouded
sources from which only neutrinos can emerge, as is the case for terrestrial beam dumps at CERN and Fermilab, are also a possibility.

\begin{figure}[h!]
\centering\leavevmode
\includegraphics[height=3.5in]{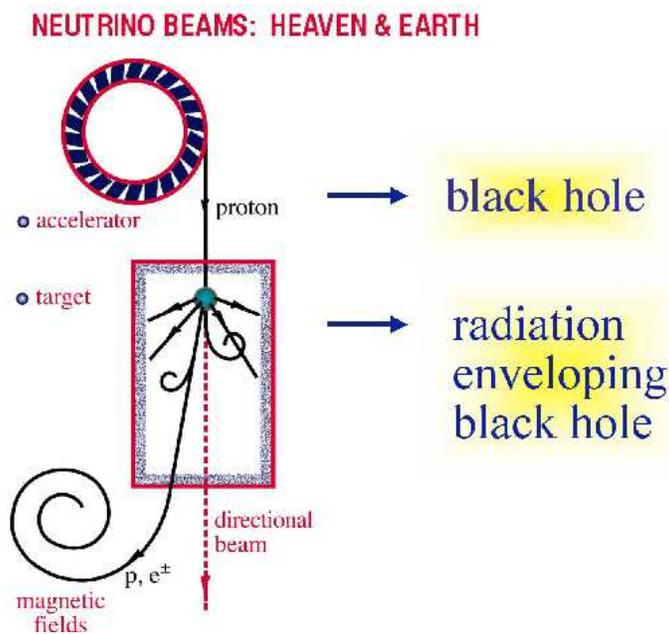}
\caption{Diagram of cosmic accelerator and beam dump.  See text for
discussion.}
\label{five}
\end{figure}

How many neutrinos are produced in association with the cosmic ray beam?
The answer to this question, among many
other\cite{snowmass},  provides the rational for building
kilometer-scale neutrino detectors.

Let's first consider the question for the accelerator beam producing
neutrino
beams at an accelerator laboratory. Here the target absorbs all parent
protons as well as the muons, electrons and gamma rays (from $\pi^0
\rightarrow \gamma + \gamma$)
produced. A pure neutrino beam exits the dump. If nature constructed
such a ``hidden source" in the heavens,
conventional astronomy has not revealed it. It cannot be the source of
the
cosmic rays, however, for which the dump must be partially transparent to
protons.

At the other extreme, the accelerated proton interacts, e.g. with photons in the radiation field surrounding the black hole, via the reaction
\begin{equation}
p + \gamma \rightarrow n + \pi^+ \mbox{ or } p + \pi^0.
\end{equation}
The $n,p$ in the final state become the observed cosmic rays and, if all particles escape without further interaction, the flux of pions and their neutrino decay products is directly related to the observed flux of cosmic rays by Eq.\,(8). The neutrino flux for such a transparent cosmic ray
source is referred to as the
Waxman-Bahcall flux \cite{wb1,wb2} and is shown as the horizontal
lines labeled ``W\&B" in
Fig.\,6. The calculation is valid for $E\simeq100$\,PeV. When evaluating the flux at both lower and higher cosmic ray energies, larger neutrino fluxes are inferred\cite{R1,R2}. This is shown as the non-horizontal line labeled
``transparent" in Fig.\,6. On
the lower side, the neutrino flux is higher because it is normalized to a
larger cosmic ray flux. On the higher side, there are more cosmic
rays in the dump to produce neutrinos because the observed flux at Earth
has been
reduced by absorption on microwave photons, the GZK-effect. The increased
values of the neutrino flux are also shown in Fig.\,6. The gamma ray
flux of $\pi^0$ origin associated with a transparent source is
qualitatively at the level of the observed flux of non-thermal TeV gamma
rays from individual sources\cite{halzenzas}.

\begin{figure}[t]
\centering\leavevmode
\includegraphics[height=2.5in]{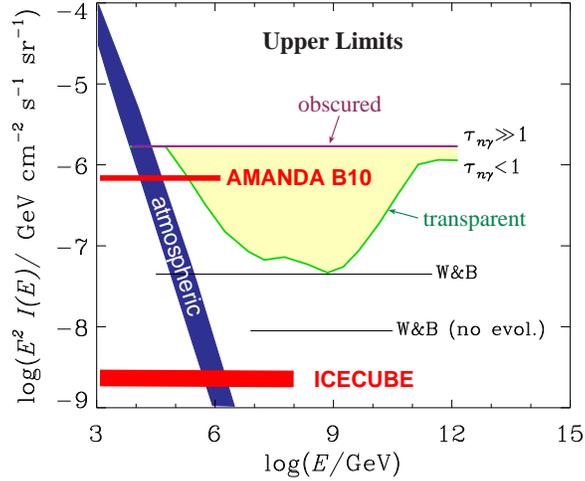}
\caption{The neutrino flux from compact astrophysical accelerators.
Shown
is the range of possible neutrino fluxes associated with the the highest
energy cosmic rays.  The lower line, labeled ``transparent", represents
a source where each cosmic
ray interacts only once before escaping the object.  The upper line,
labeled ``obscured", represents an ideal neutrino source where all
cosmic rays escape in the
form of neutrons.  Also shown is the ability of AMANDA and IceCube to
test
these models.}
\label{six}
\end{figure}

Nothing prevents us, however, from imagining heavenly beam dumps with
target densities somewhere between those of hidden and transparent
sources. When
increasing the target photon density, the proton beam is absorbed in the
dump and the number of neutrino-producing protons is enhanced relative to
those escaping the source as cosmic rays. For the extreme source of this
type, the observed cosmic rays are all decay products of neutrons
with larger mean-free paths in the dump. The flux for such a source is
shown as the upper horizontal line in Fig.\,6.

The above fluxes are derived from the requirement that theorized neutrino
sources
do not overproduce cosmic rays. Similarly, observed gamma ray fluxes
constrain potential neutrino sources because for every parent charged
pion
($\pi^{\pm}\rightarrow l^{\pm}+\nu$), a neutral pion and two gamma rays
($\pi^0\rightarrow \gamma + \gamma$) are produced. The
electromagnetic
energy associated with the
decay of neutral pions should not exceed observed astronomical fluxes.
These calculations must take into account cascading of the
electromagnetic
flux in the source and in the interstellar background photon and magnetic fields. A simple
argument relating high-energy photons and neutrinos produced by secondary
pions can still be derived by relating their total energy and allowing
for
a steeper photon flux as a result of cascading. Identifying the photon
fluxes with those of non-thermal TeV
photons emitted by supernova remnants and blazers, we predict neutrino
fluxes at the level as the Waxman-Bahcall flux\cite{gammanu}. It is
important to
realize however that there is no evidence
that TeV photons are the decay products of $\pi^0$'s. The sources
of the cosmic rays have not been revealed by photon or proton astronomy
\cite{gas1,gas2,gas3,gas4}; see however reference \cite{cangoroo}.

For neutrino detectors to succeed they must be sensitive to the range of
fluxes covered in Fig.\,6. The AMANDA detector has already entered the
region of sensitivity and is eliminating specific models which predict
the largest neutrino fluxes within the range of values allowed by general
arguments. The IceCube detector, now under construction, is sensitive to
the full range of beam dump models. IceCube will reveal the
sources of the cosmic rays or derive an upper limit that will
qualitatively raise the bar for solving the cosmic ray puzzle. The
situation could be nothing but desperate with the escape to top-down
models being cut off by the accumulating evidence that the highest energy
cosmic rays are not photons. In top-down models, decay products
eventually materialize as quarks and gluons that fragment into jets of (mostly) neutrinos, few photons and very few protons\cite{topdef}.

\section{High Energy Neutrino Telescopes}

Although neutrino telescopes have multiple interdisciplinary science
missions, the search for the sources of the highest-energy cosmic
rays stands out because it clearly identifies the size of the
detector required to do the science\cite{snr1}.

Whereas it has been realized for many decades that the science is compelling\cite{gzk1,gzk2,gzk3,gzk4}, the real challenge has been to
develop a reliable, expandable and affordable detector technology.
Suggestions to use a large volume of deep ocean water for high-energy
neutrino
astronomy were made as early as 1960. In the case
of the muon neutrino, for instance, the neutrino ($\nu_\mu$)
interacts with a hydrogen or oxygen nucleus in the water and produces
a muon travelling in nearly the same direction as the neutrino. The
blue Cerenkov light emitted along the muon's $\sim$kilometer-long
trajectory is detected by strings of photomultiplier tubes deployed
deep below the surface. Collecting muons of neutrino origin far outside the detector, the effective detector volume exceeds the volume instrumented. With the first observation of neutrinos in
the Lake Baikal and
the (under-ice) South Pole neutrino telescopes, there is optimism
that the technological challenges to build kilometer-scale neutrino telescopes can finally be met.

The first generation of neutrino telescopes, launched in 1975 by the bold
decision of the DUMAND collaboration to construct such an
instrument, are designed to reach a large telescope area
and detection volume for a neutrino threshold of order 10~GeV. The
optical requirements of the detector medium are severe. A large
absorption length is required because it determines the spacings of
the optical sensors
and, to a significant extent, the cost of the detector. A long
scattering length is needed to preserve the geometry of the Cerenkov
pattern. Nature has been kind and offered ice and water as adequate
natural Cerenkov media. Their optical properties are, in fact,
complementary. Water and ice have similar attenuation length, with
the role of scattering and absorption reversed. Optics seems, at
present, to drive the evolution of ice and
water detectors in predictable directions: towards very large
telescope area in ice exploiting the long absorption length, and
towards lower threshold and good muon track reconstruction in water
exploiting the long scattering length.

DUMAND, the pioneering project located off the coast of Hawaii,
demonstrated that muons could be detected by this
technique\cite{dumand}, but the planned detector was never realized. A
detector composed of 96 photomultiplier tubes located deep in Lake
Baikal was the first to demonstrate the detection of neutrino-induced
muons in natural water\cite{baikal2,baikal3}. In the following years,
{\it NT-200} will be operated as a neutrino telescope with an effective
area between $10^3\mbox{--}5\times 10^3$\,m$^2$, depending on energy.
Presumably too small to detect neutrinos from extraterrestrial
sources, {\it NT-200} will serve as the prototype for a larger
telescope. For instance, with 2000 OM, a threshold of 10--20\,GeV and an effective area of $5\times10^4\mbox{--}10^5$\,m$^2$, an
expanded Baikal telescope would fill the gap between present
detectors and planned larger detectors of cubic kilometer
size. Its key advantage would be low threshold.

The Baikal experiment represents a proof of concept for deep ocean
projects. These do however have the advantage of larger depth and
optically
superior water. Their challenge is to find reliable and affordable
solutions to a variety of technological challenges for deploying a
deep underwater detector. The European collaborations
ANTARES\cite{antares,antares1,antares2} and
NESTOR\cite{nestor,nestor1} have planned initial deployments of large-area detectors in the Mediterranean Sea within the next year.
The NEMO Collaboration is conducting site studies and R\&D for a
future kilometer-scale detector in the Mediterranean\cite{NEMO}.

It is however the AMANDA detector using natural Antarctic ice that has reached the $\sim10^4$\,m$^2$ telescope area envisaged by the DUMAND project a quarter of a century ago. It has operated for 3 years with 302 optical sensors and for another 3 years with 677. More than 3000 neutrinos, well separated from background, have been collected. We concentrate on the performance and first science of this detector in the rest of the lectures.

\section{The AMANDA Detector}

Since 1996 the AMANDA telescope has been taking data over Antarctic winters while construction proceeds in the summer; with 80 optical modules (OM) in 1996, with 302 during 1997-1999 (AMANDA-B) and 667 OM from 2000-present (AMANDA-II). The detectors instrument 6,000 and 16,000 kilotons of ultra-transparent ice, respectively. More than 3,000 clearly identified neutrinos have been collected thus far.

The AMANDA-B high energy neutrino telescope consists of 302 optical
modules on 10 strings.  Each OM comprises a photomultiplier
tube (PMT) with passive electronics housed in a glass pressure vessel.
The OM are deployed within a cylindrical volume 120~m in diameter
and 500~m in height at depths of 1500 to 2000~m below
the surface of the South Pole ice cap. At this depth the optical
properties of the ice are well suited for reconstructing the Cerenkov
light pattern emitted by relativistic charged
particles~\cite{ice-properties}. An electrical cable provides high voltage to the
PMT and transmits its signals to the surface data acquisition electronics.  A
light diffuser ball connected via fiber optic cable to a laser at the
surface is used for calibration purposes.  Copious down-going cosmic
ray muons, roughly 100 every second, are also used for calibration purposes.

\begin{figure}[h!]
\centering\leavevmode
\includegraphics[scale=0.55]{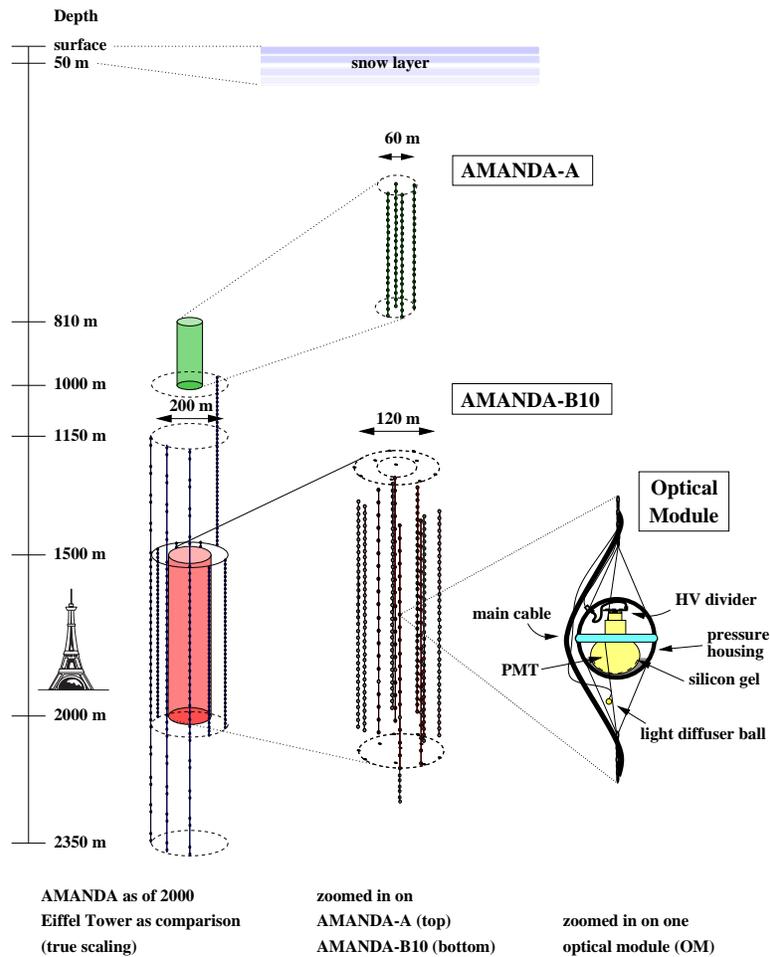}
\caption{AMANDA-B consists of 302 optical modules in a cylindrical volume with a diameter of 120~m diameter and a height of 500~m. With an additional 9 strings, AMANDA-II (on the left) consists of 667 OM in a cylindrical
volume 200~m in diameter. Both detector configurations have taken data over 3 Antarctic winters.}
\label{fig:amanda-detector}
\end{figure}

In January 2000, AMANDA-B was enlarged to a total of 19 strings with
667 OM to form AMANDA-II.  The final detector is 200~m in diameter with approximately the same height and depth as AMANDA-B.
Figure~\ref{fig:amanda-detector} shows a schematic diagram of AMANDA.

\subsection{Atmospheric Neutrinos and AMANDA Calibration}

\begin{figure}[htb]
\centering\leavevmode
\includegraphics[scale=0.75]{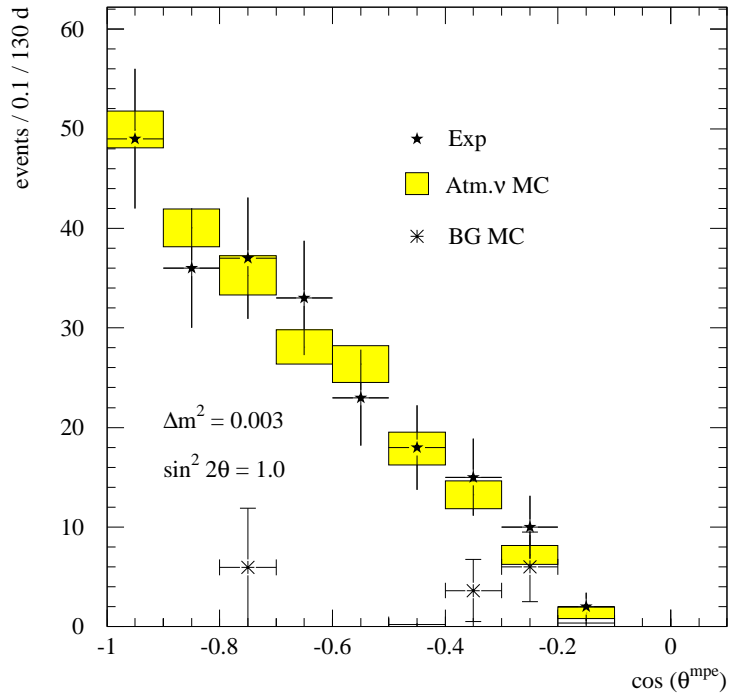}
\caption{Number of upward-going muon events in AMANDA-B data from
         the Antarctic winter 1997, as a function of zenith angle ($\cos{\theta} =
         -1.0$ is vertically up in the detector).  The data are shown
         as dots and the Monte Carlo simulation as boxes.  The simulation was performed
         with the neutrino oscillation parameters as indicated.
         The predicted signal efficiency is 4\% and the background level 10\%, with both
         numbers improving near the vertical and degrading near the horizon reflecting the shape of the partially deployed detector.
         Simulations indicate that 90\% of these events lie in an energy range 66~GeV $< E_\nu < 3.4$~TeV.}
\label{fig:b10-atmnu}
\end{figure}

AMANDA has been commissioned as a neutrino telescope in the $10,000\rm\ m^2$ class by demonstrating its ability to reconstruct upward-going muons produced by atmospheric muon neutrinos~\cite{nature,b10-atmnu}.  A fraction of the atmospheric
muon neutrinos produced in the northern hemisphere travel through the
earth, interact with the earth or the ice near AMANDA and
produce muons that can be detected and reconstructed.  Using data
collected by AMANDA-B in 1997, they have reconstructed roughly 300
upward-going muons which are, as shown in Fig.~\ref{fig:b10-atmnu}, in agreement with the predicted rate and zenith angle distribution of atmospheric neutrinos produced in the northern hemisphere. The prediction is based on the extrapolation of fluxes measured at lower energies in other underground detectors and include neutrino oscillation.

AMANDA-B data has also been used to set competitive limits on
WIMPs \cite{b10-wimps}, monopoles~\cite{b10-monopoles} reaching one order of magnitude below the Parker bound, extremely
energetic neutrinos~\cite{b10-ehe}, UHE $\nu_\mu$ point
sources~\cite{b10-point-source} and diffuse fluxes~\cite{b10-diffuse}. The detector is also sensitive to
bursts of low energy neutrinos from
supernovae~\cite{b10-supernova}.

\begin{figure}[h!]
\centering\leavevmode
\epsfig{file=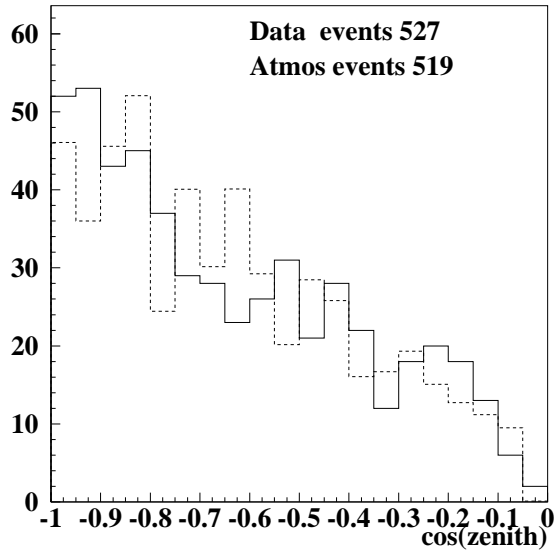,width=3.25in}
\caption{Number of upward-going muon events in AMANDA-II data from the
         year 2000 as a function of zenith angle, using a preliminary
         set of selection criteria.  There are a total of 527 events
         in the data or roughly 4 per day (solid line), while 519 events are predicted by the
         atmospheric neutrino Monte Carlo (dashed line).  Simulations
         indicate that these events have an energy of $\rm 100~GeV < E_\nu < 1~TeV$.  With more sophisticated selection criteria one expects larger event rates and improved response near the horizon.}
\label{fig:aii-atmnu}
\end{figure}

A preliminary analysis of atmospheric neutrino data taken with\break
AMANDA-II in 2000 demonstrates the substantially increased effective volume of the
enlarged detector.  Compared to the analysis using AMANDA-B data,
fewer selection criteria are required to extract a larger and
qualitatively cleaner sample of atmospheric neutrino-induced muons. Because of the simplified and robust analysis, neutrino separation from background is now possible in real time and has been implemented in 2002. Figure~\ref{fig:aii-atmnu} shows the excellent agreement between
data and simulation achieved with a preliminary set of selection
criteria applied. Note the improved
angular response close to the horizon. With more sophisticated selection criteria one
expects to extract at least 3 times more neutrino events in AMANDA-II relative to
AMANDA-B for equivalent live-times.

\subsection{Search for Cascades with AMANDA-B and -II}

1997 AMANDA data has also been used to perform a search for the Cerenkov light produced by electromagnetic or hadronic showers ({\it
cascades}) induced by high-energy extraterrestrial
neutrinos. Demonstrating cascade sensitivity is an important step
for neutrino astronomy because cascades probe all neutrino
flavors, whereas the muon channel is sensitive to $\nu_\mu$ only. Electron neutrinos produce cascades via the charged current
interaction and all neutrino flavors produce cascades via the neutral
current interaction. Cascade-like events are also produced in charged
current $\nu_\tau$ interactions. Compared to muons, cascades provide more accurate energy measurement, superior separation from background, but worse angular resolution. Cascades have a superior energy resolution because all energy is
deposited in a small volume of ice of order 10\,m diameter and because the number of Cerenkov photons scales linearly with the deposited energy. As with muons, cascades are easier to
identify and reconstruct as the instrumented detector volume increases. After application of simple selection criteria that reduce the down-going muon background while preserving potential signal events, vertex position, energy and direction of the cascade are
reconstructed using maximum likelihood methods. These take
into account the expected Cerenkov pattern after absorption and scattering of the light in the ice~\cite{picrc,amanda-cascades}.

In the absence of a tagged source of high energy neutrino-induced
cascades, one determines the response of the detector using {\it in-situ} light sources. The successful
reconstruction of data taken with a pulsed laser and the reconstruction of isolated showers produced by the catastrophic energy loss of muons, have demonstrated that the detector is sensitive to high energy cascades~\cite{amanda-cascades}.

The 90\% C.L. limit on a diffuse flux of $\nu_e+\nu_\mu+\nu_\tau+
\overline{\nu}_e+\overline{\nu}_\mu+\overline{\nu}_\tau$ for
neutrino energies between 5~TeV and 300~TeV is:
\begin{equation}
   \label{nu_l_limit_eq}
   E^2\frac{d\Phi}{dE} < 9.8 \times 10^{-6} \; \mathrm{GeV\,cm^
{-2}\,s^{-1}\,sr^{-1}},
\end{equation}
assuming an oscillated neutrino flux with 1:1:1 flavor composition. The 90\% C.L. limit on the diffuse flux of $\nu_e+\overline{\nu}_e$ for
neutrino energies between 5~TeV and 300~TeV
is:
\begin{equation}
   \label{nu_e_limit_eq}
   E^2\frac{d\Phi}{dE} < 6.5 \times 10^{-6} \;
   \mathrm{GeV\,cm^{-2}\,s^{-1}\,sr^{-1}}.
\end{equation}
Note that since the limit in Eq.~(\ref{nu_l_limit_eq}) is on the sum of the
fluxes of all neutrino flavors and the limit in
Eq.~(\ref{nu_e_limit_eq}) is on an individual flavor, the former
should be divided by three when compared to the
latter.

Data from AMANDA-II is currently under study and, as with the
analysis of atmospheric neutrinos, preliminary results demonstrate the enhanced power of the larger AMANDA-II
detector.  Angular acceptance improves to nearly $4\pi$, backgrounds
are much easier to reject, and energy acceptance improves by a factor
of three to $E_\nu \sim 1$~PeV.  In accordance with blind analysis procedures, 20\% of the AMANDA-II data from the year 2000 has been analyzed resulting in a preliminary limit that is already lower than the one above by a factor of 3.  For a source with a UHE neutrino flux at the
current best limit~\cite{b10-diffuse}, they expect eight UHE-neutrino-induced cascade events in the full 2000 dataset on
an expected background of less than one event.

\subsection{Search for UHE $\nu_\mu$ from Point Sources}

The AMANDA collaboration has performed a general search for  the continuous emission of muon
neutrinos from a spatially localized direction in the northern sky.
Backgrounds are reduced by requiring a statistically significant
enhancement in the number of reconstructed upward-going muons in a
small bin in solid angle. The background for a
particular bin can be calculated from the data by averaging over the
data outside the bin in the same declination band.  In
contrast to the searches previously discussed, this search is more tolerant of the
presence of background and the signal is therefore optimized on $S/\sqrt{B}$,
where $S$ represents the signal and $B$ the background, rather than on
$S/B$, which emphasizes signal purity.

The analysis of AMANDA-B data taken in 1997 has been reported in~\cite{b10-point-source}.  With AMANDA-II, one achieves larger effective area and improved sensitivity to events near the horizon because the detector has double the number of PMT and a larger lever arm in
the horizontal dimension.  Assuming a customary $E^{-2}$ source spectrum, and a flux of
$10^{-8}\rm\,GeV\,cm^{-2}\,s^{-1}\,sr^{-1}$, one predicts two
signal and one background event in a $6^\circ \times 6^\circ$ angular
bin.  This flux represents an interesting benchmark because, at this level, AMANDA should observe TeV sources provided the number of gamma rays and neutrinos are roughly equal as expected from cosmic ray accelerators producing pions. Preliminary sensitivities to a sample of point sources are given in
Table~\ref{table:point_source_sensitivities}.  In order to achieve
blindness in this analysis the right ascension of each event (i.e., its azimuthal
angle) has been scrambled. At the South Pole this effectively
scrambles the event time. The data will only be unscrambled after final selection criteria have been set.

\begin{table}[htb]
\caption{Example of AMANDA-II sensitivity to point sources in data taken in 2000.  The sensitivity
         is defined as the predicted average limit from an
         ensemble of experiments with no signal, and is calculated using
         background levels predicted from off-source data.}
\label{table:point_source_sensitivities}
\renewcommand{\arraystretch}{1.2} %
\centering\leavevmode
\begin{tabular}{lr@{\qquad}r@{\qquad\qquad}r@{\qquad\qquad}} \hline
Source              & \multicolumn{1}{c}{Declination}
                            & \multicolumn{1}{c}{$\mu$ ($\times 10^
{-15}$cm$^{-2}$s$^{-1}$)}
                                   & \multicolumn{1}{c}{$\nu$ ($\times 10^{-8}$cm$^{-2}$s$^{-1}$)}\\ \hline
SS433               & 5.0   & 11.0 & 2.4 \\
Crab                & 22.0  & 4.0  & 1.3 \\
Markarian 501       & 39.8  & 2.5  & 1.0 \\
Cygnus X-3          & 41.5  & 2.6  & 1.1 \\
Cass. A             & 58.8  & 2.1  & 1.0 \\  \hline
\end{tabular}
\end{table}

\subsection{Search for a Diffuse Flux of $\nu_\mu$ Neutrinos}

The search for diffuse sources of high energy $\nu_\mu$--induced muons is closely associated with the analysis isolating atmospheric $\nu_\mu$--induced
muons as both analyses require a sample of muon tracks well separated from
misreconstructed downward-going atmospheric muons.
Since high-energy muons deposit more energy in the
detector volume than low-energy muons, one isolates high energy events by requiring a high channel density, $\rho_{\rm ch} > 3$,
where the channel density is defined as the number of hit channels per
10~m tracklength.  The background in the signal region is estimated
by extrapolation of lower-energy data satisfying $\rho_{\rm ch} < 3$.

Using a 20\% sample of the year 2000 AMANDA-II data,  6 events satisfying all selection criteria have been identified.  Simulations for a $E^{-2}$ power law
spectrum predict 3.0 events from a UHE neutrino flux at the current
best limit~\cite{b10-diffuse} and 1.9 events from atmospheric neutrino
interactions. Again a subsample of the data is used in order to
achieve blindness in this analysis. The predicted average limit from an ensemble of experiments with no
signal, or {\it sensitivity}, is roughly $1.3 \times
10^{-6}$~GeV$\,$cm$^{-2}$s$^{-1}$sr$^{-1}$, and the preliminary limit is less
than roughly $10^{-6}$~GeV$\,$cm$^{-2}$s$^{-1}$sr$^{-1}$.  With only 20\% of the year 2000 data, AMANDA-II matches the limit obtained with the {\it full} sample of AMANDA-B
data from 1997.

\subsection{Search for $\nu_\mu$ from GRB}

The search for $\nu_\mu$--induced muons from gamma-ray bursts (GRB) leverages temporal and directional information from satellite observations to realize a nearly background-free analysis. The steady 100\,Hz cosmic ray muon background is indeed limited during the short lifetime and in the direction of a specific GRB. Assuming the broken power-law spectrum predicted by the standard fireball model\cite{grb}, AMANDA has searched for muon neutrinos in coincidence with GRB. Off-source and off-time data are used to estimate background and to achieve blindness in the analysis. Data spanning the years 1997--2000 have been analyzed with roughly 500 GRB in coincidence with satellite experiments.

Detector stability over 10~s periods, a typical signal window, is an important measure
of how effective this analysis can be.  They have shown that the counting rate per 10~s bin in a time window of $\pm 1$~hour around a GRB is Gaussian, revealing no instrumental effects that can mimic GRB.  A limit has been obtained that reaches to within an order of magnitude of the typical event rate of 20 per kilometer square per year anticipated by fireball phenomenology\cite{amandagrb}. With two more years of data on tape, discovery is possible, especially because the signal is mostly generated by favorable fluctuations in distance and energy of individual GRB\cite{grb}.

Overall, AMANDA represents a proof of concept for the construction of a kilometer-scale neutrino observatory, IceCube.

\section{IceCube: a Kilometer-Scale Neutrino Observatory}

IceCube is an instrument optimized to detect and characterize neutrinos of all flavors from sub-TeV to the highest energies; see Fig.\,10. It will consist of 80 kilometer-length strings, each instrumented with 60 10-inch photomultipliers spaced by 17~m.
The deepest module is 2.4~km below the surface. The strings are
arranged at the apexes of equilateral triangles 125\,m on a side. The instrumented (not effective!) detector volume is a cubic kilometer. A surface air shower detector, IceTop, consisting of 160 Auger-style Cerenkov detectors deployed over 1\,km$^{2}$ above IceCube, augments the deep ice component by providing
a tool for calibration, background rejection and air shower physics.

\begin{figure}[h]
\centering\leavevmode
\includegraphics[width=2.7in]{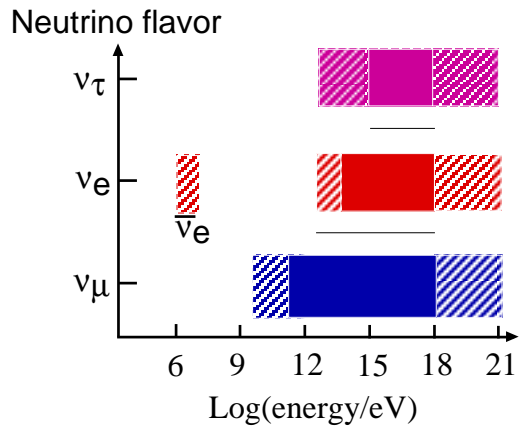}
\caption{Although IceCube detects neutrinos of all energies and flavor above a
threshold of $\sim 0.1$\,TeV, it can identify their flavor and measure
their energy only in the ranges shown.}
\end{figure}

IceCube will offer great advantages
over AMANDA-II beyond its larger size: it will have a higher
efficiency and superior angular resolution in reconstructing tracks, map showers from electron- and
tau-neutrinos (events where both the production and decay of a $\tau$
produced by a $\nu_{\tau}$ can be identified) and, most
importantly, measure neutrino energy. Simulations, backed by AMANDA data, indicate that the
direction of muons can be determined with sub-degree accuracy and
their energy measured to better than 30\% in the logarithm of the
energy. The direction of showers will be reconstructed to better
than 10$^\circ$ above 10\,TeV and the response in energy is linear and better than 20\%. Energy resolution is critical because,
once one establishes that the energy exceeds 1\,PeV, there is no
atmospheric muon or neutrino background in a kilometer-square detector and full sky coverage is achieved.

At lower energies the backgrounds are down-going
cosmic ray muons, atmospheric neutrinos, and the dark noise
signals produced by the photomultipliers themselves.  The simulated trigger  rate of
down-going cosmic ray muons in IceCube is 1700\,Hz while the rate
of atmospheric neutrinos ($\nu_\mu$ and $\overline{\nu}_\mu$) at trigger
level is 300 per day.  Depending on the type of signal to be searched for,
this background is rejected using direction, energy, and neutrino flavor.
At energies below 1\,PeV, neutrino astronomy must focus on upward going
neutrinos.  At energies above 1 PeV, the cosmic ray muon background
disappears while the low energy cosmic ray background can be rejected
using an energy rejection cut.  The background contribution
from dark noise is not significant because the total dark noise rate of the optical sensor
in situ is expected to be less than 0.5\,kHz.

The highest sensitivity for astrophysical point sources can be achieved
with muons. The muon channel stands out for two reasons: i) muons allow a sub-degree angular resolution over a wide energy range and ii) the effective volume for muons exceeds the geometric
volume of the detector by factors of 10 to more than 50 depending on
energy because of the muon range that is of order kilometers.
Due to the long range of high energy muons the interaction of the $\nu_\mu$
can be detected at distances of tens of kilometers outside the detector.
We will here exclusively focus on the muon detection because it provides the benchmark
sensitivity for some of the fundamental goals of high energy neutrino
astronomy.

\subsection {Detector Design}

The detector consists of 4800 photomultipliers instrumenting a volume of 1\,km$^3$. The transmission of analogue PMT signals to the surface, used in AMANDA, has been abandoned. The photomultiplier signals will be captured and digitized inside the OM.  The digitized
signals are given a global time stamp with a precision of $<$5\,ns and
transmitted to the surface.  The digital messages are sent to a string
processor, a global event trigger and an event builder.  All time calibrations
will be automated.  The geometry of the detector will be measured to a
precision of better than 2\,m during deployment.  It will be
calibrated more precisely ($<1$\,m) with light flashers on-board the OM and also with cosmic ray muons.  Both methods have been tested and successfully applied in
AMANDA. High energy signals and complex events can be calibrated with
powerful lasers deployed with the detector.  Its absolute orientation can
be determined from coincidences with the surface air shower array and by observation of the
shadow of the moon, not possible with smaller detectors such as AMANDA.  Once events are built in the surface DAQ, data
will be
processed and filtered. A reduced data set will be sent on a daily basis to the Northern
hemisphere for further processing and data analysis.
Twenty four
hour satellite connectivity will be available for important messages.
Construction of the detector is expected to commence in the Austral summer
of 2004/2005 and continue for 6 years, possibly less.  The growing detector will take data during construction, with each string coming online within days of deployment.

\subsection{Performance}
\label{sec:performance}

\subsubsection{Telescope Area}

In detailed simulations based on AMANDA data, the response of the detector has been evaluated to cosmic
ray muons, to atmospheric neutrinos and to a hypothetical $E^{-2}$ cosmic neutrino
spectrum generated by a generic shock acceleration mechanism
\cite{performance}.  The event rates, normalized to one year of on-time, are
listed in Table~\ref{tab:passclasses}. Listed are event rates at trigger level as well as for full event reconstruction, with cuts applied for the
rejection of the cosmic ray muon background. This is referred to as ``level\,2''. The trigger consists of a minimum of 5 local coincidences; a local coincidence is between two adjacent PMT or a pair separated by at most one PMT. Background from misreconstructed down-going cosmic ray muons can be rejected by a straightforward cut in energy. For a cosmic neutrino flux of $E_{\nu}^{2} \times dN_{\nu}/dE_{\nu} =10^{-7}\rm\,GeV\,(cm^2\, s\, sr)^{-1}$, roughly the final sensitivity of AMANDA-II\cite{b10-diffuse}, more
than 1000 signal events per year are predicted. At this stage, the background of $10^{5}$ events, consists mostly of atmospheric neutrinos\cite{lipariatm} and prompt muons from charm decay in the atmosphere\cite{rqpm}.

\begin{table}[htb]
\renewcommand{\arraystretch}{1.2}
\tabcolsep=2em
  \begin{center}
 \caption[]{Event rates are given for 1 year of signal and background.
The signal assumes a cosmic neutrino flux of $ E^{2}_{\nu} \times dN_{\nu}/dE_{\nu} = 10^{-7}$\,\diffunit, roughly the ultimate sensitivity of AMANDA-II.
The calculation of atmospheric neutrino induced muon events
is based on \cite{lipariatm} and it includes the
prompt component according to~\cite{rqpm}. }
\label{tab:passclasses}
   \begin{tabular}{ccc}
    \hline
          &  Trigger                                       &   Level 2    \\
    \hline
     Cosmic $\nu$    &  $3.3\times 10^{3}$      &  $1.1 \times 10^{3}$ \\
  Atm $\nu$    &  $8.2  \times 10^{5}$                &  $9.6  \times 10^{4}$    \\
     Atm $\mu$ & $4.1 \times 10^{10}$  &  $10  \times 10^{4}$       \\
     \hline
    \end{tabular}
   \end{center}
\end{table}

The usual way to characterize the performance of the detector is to evaluate the ``effective detector area'' which is defined as
\begin{equation}
    A_{\mathrm{eff}}(E_{\mu},\Theta_{\mu} ) =
  \frac{N_{\mathrm{detected}}(E_{\mu}, \Theta_{\mu})}{N_{\mathrm{generated}}(E_{\mu},\Theta_{\mu} )} \times A_{\mathrm{gen}}.
\end{equation}
$N_{\mathrm{generated}}$ is the number of muons in the test sample with energy
$E_\mu$ and incident zenith angle $\Theta_{\mu}$. The energy of the muon is defined at the point of closest approach to the center of the detector. $N_{\mathrm{detected}}$ is
the number of events that actually trigger the detector, or pass the cut level
under consideration.

\begin{figure}[htp]
\begin{center}
\mbox{}
\epsfig{file=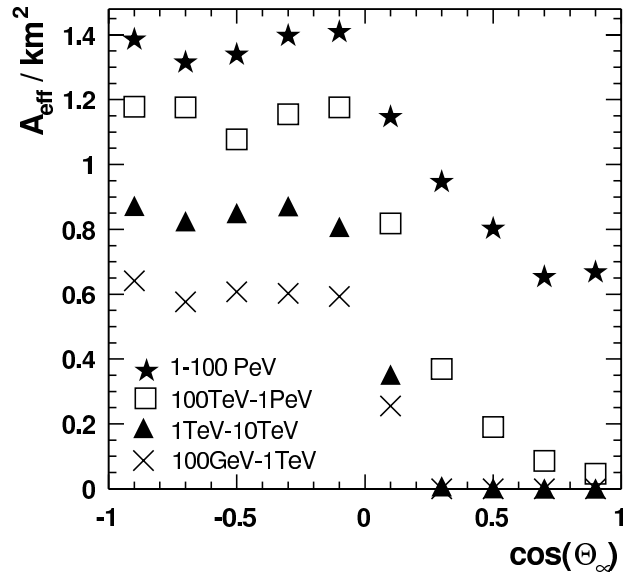,width=.65\textwidth}
\end{center}

\caption{The effective area for neutrino-induced muons is shown as a function of the zenith angle after applying level\,2 cuts.
\label{fig:aeff.en}}
\end{figure}

Figure\,\ref{fig:aeff.en} shows the effective area for four energy intervals
as a function of the zenith angle of the incident muon
track. The effective trigger area reaches one square kilometer at an energy of a
few hundred GeV. Roughly 50\% of all triggered events pass the ``standard
selection'' (level\,2), independent of the muon energy. The detector reaches an effective detection area of one
square kilometer for upward moving muons in the TeV range.  Importantly, above 100\,TeV
the selection allows the detection of down-going neutrinos for observation of the southern
hemisphere ($\cos \theta > 0$). In the PeV range the effective area
for down-going muons is above 0.6 km$^{2}$, increasing towards the horizon.
This means that IceCube can be operated as a full sky observatory for neutrinos with energy in excess of $\sim 1$~PeV.

\subsubsection{Angular Resolution}

\begin{figure}[htp]
\begin{center}
\epsfig{file=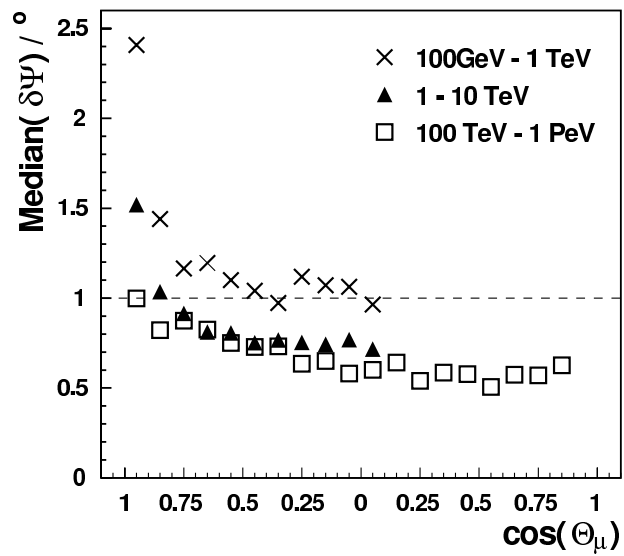,width=.65\textwidth}
\end{center}

\caption{ Pointing resolution for muons.
Shown is the median space angle error of the reconstructed direction
as a function
of  the zenith angle of the incident track.
\label{fig:pointing}
}
\end{figure}

The angular resolution of the detector is relevant for the search
for  neutrinos from point sources. Exploiting the angular resolution of the
detector,
background events can be more easily eliminated by restricting the search
to a small angular region around the known direction of the object under
investigation. Using AMANDA techniques only, an angular resolution approaching $0.5^\circ$ is achieved; see Fig.\ref{fig:pointing}. Significant improvement of this resolution is expected with the development of reconstruction algorithms that include amplitude and
waveform information not provided by AMANDA detection methods.

\subsubsection{Sensitivity to Astronomical Neutrino Fluxes}

\noindent{\it Diffuse Fluxes}.
As previously discussed, theoretical models of astrophysical fluxes of high energy neutrinos are often linked to the known flux of the very high energy cosmic rays of extragalactic origin; see Fig.\,6. The harder energy spectrum, typically $E^{-2}$, expected for an astrophysical neutrino flux
can be used to discriminate against the softer atmospheric
neutrino background. The number of
optical modules in an event that detect at least one photon, called the channel
multiplicity $N_{\mathrm{ch}}$, is used as a very simple and robust energy observable. Atmospheric neutrino-induced events have typical
channel multiplicities of 30 to 60 channels.
The assumed high energy signal dominates the atmospheric neutrino
background above channel multiplicities in the vicinity of 200. For an optimized energy cut $N_{\mathrm{ch}}>227$, a simulated source with spectrum $E_{\nu}^{2} \times dN_{\nu}/dE_{\nu} = 1
\times 10^{-7}\rm\,GeV\,(cm^2\, s\, sr)^{-1}$, previously introduced, results in 74 signal
events in one year of operation compared to a background of 8 atmospheric neutrinos. The background was calculated including prompt decays of charmed mesons which may contribute as much as 80\% to the final atmospheric sample following reference\cite{rqpm}. After three years of operation an overall flux limit of
 $E^2 \times dN_{\nu}/dE_{\nu} = 8.1 \times 10^{-9}\rm\,GeV\,(cm^2\, s\, sr)^{-1}$ is
 obtained. This is more than
two orders of magnitude below the current limit obtained with AMANDA. The energy cut
results in a detection threshold of 200\,TeV.
The sensitivity  obtained after one year of data taking
is well below the Waxman-Bahcall flux previously discussed.

\noindent{\it Sensitivity to Point Sources}.
With greatly improved angular resolution, IceCube can search for point neutrino sources with a search window of one degree radius thus greatly reducing the background. The remaining small number of background atmospheric neutrinos in the search bin is rejected by a soft energy cut \Nch${}>30$. The cut eliminates atmospheric neutrinos of energies below 1\,TeV allowing for an essentially background free
detection.
With these parameters an average flux upper limit of
$E_{\nu}^2 \times  dN_{\nu}/dE_{\nu} = 5.5 \times 10^{-9}$\,\pointunit\ is obtained after one year of data taking. After five years of operation the sensitivity level will reach
$E^{2} \times dN_{\nu}/dE_{\nu} \sim 1.7 \times 10^{-9}$\,\pointunit.

\noindent{\it Gamma Ray Burst Sensitivity}.
As discussed in the context of AMANDA, the short duration of gamma ray bursts allows for an essentially background-free search. The AMANDA analysis only covers GRB in the northern sky; with IceCube a full sky search is possible at high energy. For a rate of 500 bursts over $2\pi$\,sr and 1 year, 13 neutrino induced up-going muons are predicted on a background of 0.1. This rate corresponds to the standard fireball model normalized by the assumption that GRB are the sources of the highest energy cosmic rays.

\subsubsection{Cascades and Very High Energy Neutrinos}

Due to its large and uniformly instrumented volume, IceCube also has excellent sensitivity to cascades generated by
 $\nu_e$, $\overline{\nu}_e$, $\nu_\tau$, and
$\overline{\nu}_\tau$. At energies above a few tens of TeV the detector becomes fully efficient
to cascade detection with an effective volume comparable to the
geometric volume of 1\,km$^{3}$; see Fig.~13.
At energies above 1 PeV tau events can be separated from $\nu_e$-induced cascades using a variety of methods. For extremely energetic muons, effective areas well beyond the
geometric area are achieved. Effective
areas at 10$^{18}$\,eV are estimated to reach more than 2.5\,km$^{2}$.
Reconstruction algorithms for the more complex event topologies
such as from tau events and for extreme energies beyond 10$^{18}$\,eV
are not fully developed yet. However, the high resolution and large dynamic range of the IceCube data acquisition system will provide rich information to
extract the physics from such events.

\begin{figure}[htb]
\centering\leavevmode
\epsfig{file=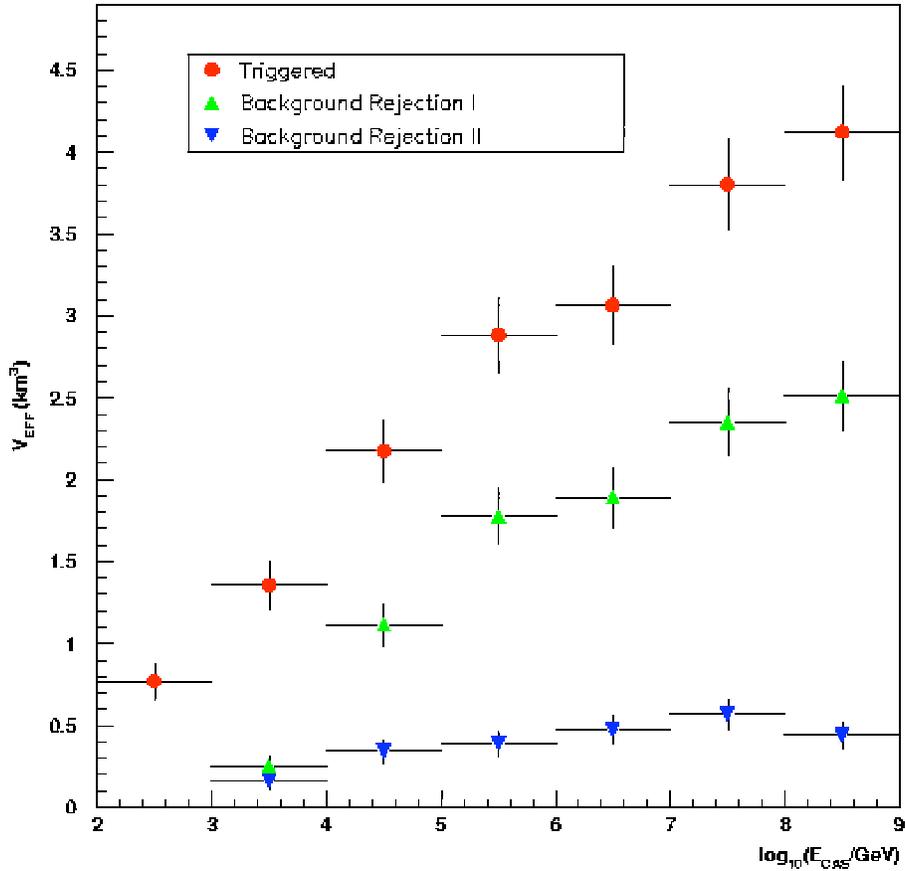,width=5.25in}
\caption{Effective volume for neutrino-induced cascades at trigger level (circles), after background suppression cuts (upward triangles), and both background cuts and requiring that the cascade falls within the volume of instrumented ice (downward triangles).}
\end{figure}

\subsubsection{Other Science Opportunities}

\noindent{\it Dark Matter}.
If Weakly Interacting Massive Particles (WIMPs) are the dark matter of
the Universe, they populate the galactic halo of our Galaxy. Some have been captured over astronomical times by the Earth and the sun where they annihilate
pairwise, producing high-energy muon neutrinos that can be searched for by
neutrino telescopes.  A favorite WIMP candidate is the lightest neutralino of the Minimal Supersymmetric Model. The energy of the neutrino-induced muons is of order $25\%$
of the neutralino mass.  The predicted muon rates from WIMPs annihilating
in the sun can be as large as $10^4\rm\, km^{-2}$ per year at $100$\,GeV and more
than a $10^{4}\rm\,km^{-2}$ up to
energies of $1$\,TeV. Current limits eliminate fluxes larger than several thousand events at energies above 100\,GeV. Simulations indicate
that IceCube should reach sensitivities below 50 muon events per year for
WIMPs from the sun \cite{edsjoe}.  Thus IceCube will play a complementary
role to future direct detection experiments like GENIUS for
annihilation in the Earth, and be competitive with such next-generation direct
detection experiments for annihilation in
the Sun.

\noindent{\it Cosmic Rays and Air Showers}.
Combined with the 1\,km$^2$ surface detector, IceCube can do unique
coincidence and anti-coincidence measurements of high energy air showers.
In addition to providing a sample of events for calibration and for study
of air-shower-induced backgrounds in IceCube, the surface array will also act as
a partial veto.  All events generated by showers with $E > 10^{15}$\,eV can
be vetoed when the shower passes through the surface array.  In addition,
higher energy events, which are a potential source of background for
neutrino-induced cascades, can be vetoed even for showers passing a long
distance outside the array.
The IceCube--IceTop coincidence data will cover an energy range starting below
the knee of the cosmic-ray spectrum to $>10^{18}$\,eV. Each event will
contain a measure of the shower size at the surface and a signal from the
deep detector produced by muons with $E > 300$\,GeV at production.  At the
high elevation of the South Pole, showers will be observed near shower maximum so
that measured shower size provides a good measure of the
total size and the energy.
The combined measurement of the muon-induced signal in IceCube
 and the shower size at the surface will yield information on the primary composition over three orders of magnitude in energy.
In particular, if the knee is due to a steepening of the rigidity spectrum,
a steepening of the spectrum of protons around $3\times 10^{15}$\,eV should
be followed by a break in the spectrum of iron at $8\times 10^{16}\rm\, eV$.  The
method for measuring composition has been developed and applied
successfully with AMANDA and the surface air shower array SPASE-2
\cite{kath-thesis}.

\subsubsection{ Supernova Detection }

Although the MeV energies of supernova neutrinos are far below the\break
 AMANDA/IceCube
trigger threshold, a supernova can be detected by observing higher
counting rates of individual
PMT over a time window of 5--10 seconds. The enhancement in rate of a single PMT will be buried
in its dark noise. However, by summing the signals from all PMT over 10~s, significant excesses are observed. With background rates more than 10 times lower than
ocean experiments, IceCube has the potential to see a supernova out to the LMC and to generate
an alarm signal.

\section{The Future Is Now}

At this point in time, several new instruments such as HiRes, the
partially deployed Auger array, Hess and Magic, Milagro, Baikal and
AMANDA-II are taking data. With rapidly
growing observational capabilities, one can express the realistic hope
that the cosmic ray puzzle will be solved soon. The solution is likely to reveal unexpected astrophysics, if not particle physics.

\section*{Acknowledgments}
This work was supported in part by DOE grant No. DE-FG02-95ER40896 and by the U.S.  National
Science Foundation, Office of Polar Programs; U.S. National Science
Foundation, Physics Division and the University of Wisconsin Alumni Research
Foundation.

\end{document}